\documentclass[11pt,a4paper]{article}
\usepackage{amssymb}
\usepackage{epsfig}
\usepackage{anysize}

\newcommand{\eff}{\mathrm{eff}}
\newcommand{\nn}{\mathrm{m}}

\begin{document}
\begin{center}
{\LARGE Effect of many-body interactions on the solid-liquid 
phase-behavior of charge-stabilized colloidal suspensions}\\
\vspace*{0.5cm}

J. Dobnikar$^{a}$\footnote{Jure.Dobnikar@uni-konstanz.de},
  R. Rzehak$^{b}$ and H.~H.~von~Gr\"unberg$^{a}$ \\
\vspace*{0.5cm}
 
{\small $^{a}$Universit\"at Konstanz, Fachbereich Physik, D-78457 Konstanz, Germany\\
$^{b}$ IFF, Forschungszentrum J\"ulich, D-52425 J\"ulich, Germany}\\
\vspace*{0.5cm}
\end{center}

\begin{abstract}
  The solid-liquid phase-diagram of charge-stabilized colloidal
  suspensions is calculated using a technique that combines a
  continuous Poisson-Boltzmann description for the microscopic
  electrolyte ions with a molecular-dynamics simulation for the
  macroionic colloidal spheres. While correlations between the
  microions are neglected in this approach, many-body interactions
  between the colloids are fully included. The solid-liquid transition
  is determined at a high colloid volume fraction where many-body
  interactions are expected to be strong.  With a view to the
  Derjaguin-Landau-Verwey-Overbeek theory predicting that colloids
  interact via Yukawa pair-potentials, we compare our results with the
  phase diagram of a simple Yukawa liquid. Good agreement is found at
  high salt conditions, while at low ionic strength considerable
  deviations are observed.  By calculating effective colloid-colloid
  pair-interactions it is demonstrated that these differences are due
  to many-body interactions.  We suggest a density-dependent
  pair-potential in the form of a truncated Yukawa potential, and show
  that it offers a considerably improved description of the
  solid-liquid phase-behavior of concentrated colloidal suspensions.\\
\end{abstract}

The classic Derjaguin-Landau-Verwey-Overbeek (DLVO) theory
\cite{verwey} predicts that an isolated pair of charged colloidal
spheres in an aqueous salt solution interacts via a repulsive Yukawa
potential at large separations. Direct experimental measurements have
confirmed the general validity of this prediction
\cite{crocker352,vondermasson1351,crocker1897}. Though originally
being designed only for pairs of colloids in isolation, it is common
practice nowadays to apply the DLVO theory also to the much more
complex case of colloids in concentrated suspensions, the situation
which occurs in most technical applications.  As we show in this
letter, treating a colloidal suspension as a system of pairwise
interacting Yukawa particles is valid only in a limited parameter
regime.  To this end a full Poisson-Boltzmann (PB) mean-field
description of the colloidal interactions has been combined with a
molecular-dynamics (MD) simulation, similar to
Ref.~\cite{fushiki6700}, to calculate the solid-liquid phase-behavior
of charge-stabilized colloidal suspensions.  While correlations
between the microions are neglected in this approach, many-body
interactions between the colloids, mediated by the screening ionic
fluid between and around the colloids, are fully included.  By
calculating the effective DLVO parameters for the same system, it is
seen that at large volume fraction of colloids and low salt
concentration, the description in terms of pair-wise interacting
Yukawa particles fails dramatically.

An estimate whether the DLVO pair potential is a good approximation in
concentrated colloidal suspensions can be based on the ratio between
i) the mean colloid-colloid distance, $d_{\nn} = \rho^{-1/3}$, in
suspensions with a density of colloidal particles $\rho$ and ii) the
inverse Debye screening length $\kappa^{-1}$, which measures the
thickness of the spherical double-layer around a single isolated
colloidal sphere due to screening by the microions.  If two or more
such spherical double-layers overlap, the screening of the colloidal
charges becomes incomplete and the charge distributions on the
colloids involved begin to interact.  If the colloid density is low,
i.e. if $d_{\nn}\kappa \gg 1$, it is obvious that such an overlap will
almost always occur for pairs of colloids only. At high colloid
densities, however, i.e. if $d_{\nn} \kappa \sim 1$, there is a high
probability that more than one other colloid is within the range of
the double-layer $\kappa^{-1}$ around any colloidal particle.  In this
case many-body forces between the colloids become important, and the
picture of pair-wise interacting particles ceases to be valid.

This scenario has been qualitatively confirmed by recent measurements
of the effective colloidal pair-interaction potential as a function of
the colloid density in 2D colloidal systems under low-salt conditions
\cite{brunner,klein}. In this context, many-body interactions show up
as a density-dependence of the effective pair-potential
\cite{ArdLouis}. Indeed, at low density, the measured pair-potential
was Yukawa-like as predicted by DLVO theory, but at high density it
showed clear deviations from a Yukawa-like behavior, particularly for
inter-colloidal distances $r \gtrsim d_{\nn}$ where the interaction
decayed to zero much more rapidly than predicted by DLVO theory.  This
observed density-dependent ''cut-off'' of the interaction potential at
the mean colloid-colloid distance $d_{\nn}$ is a direct manifestation
of many-body interactions in the system \cite{brunner}. A simple
physical explanation for this ''cut-off'' is based on the mechanism of
macro-ion shielding \cite{klein}: the interaction between two colloids
separated by a distance $r>d_{\nn}$ is likely to be screened by a
third colloid located somewhere between them.  This screening effect
of the macroionic charges, which is neglected in the DLVO theory,
qualitatively explains the observed deviations from the
Yukawa-potential.  Motivated by these experiments, we here address the
question to what extent many-body interactions affect
\emph{macroscopic} properties of {\em 3D} colloidal suspensions,
specifically, the solid-liquid phase behavior.

Colloidal crystals near melting are simulated using a technique
suggested by Fushiki \cite{fushiki6700} which combines a PB field
description for the microions with a MD simulation for the macroionic
colloids \cite{technical_paper}. Specifically, we consider a fluid of
highly charged identical colloidal spheres suspended in a
structureless medium of dielectric constant $\varepsilon$ at
temperature $T$.  The Bjerrum length characterizing the medium is
defined as $\lambda_{B} = e^{2}\beta/\varepsilon$, with $\beta =
1/kT$, $k$ Boltzmann's constant, and $e$ the elementary charge.  The
colloidal spheres have radius $a$, charge $-Ze$, and thus a surface
charge density $-e\sigma = -Ze/4 \pi a^{2}$.  They are placed in a
cubic box with periodic boundaries and the positions of their centers
are denoted by ${\bf R}_i$ ($i=1\ldots N)$.  The suspension is assumed
to be in osmotic equilibrium with an electroneutral reservoir of
monovalent point-like salt ions with total particle density $2c_{s}$
and inverse Debye screening length $\kappa=(8\pi\lambda_B
c_{s})^{1/2}$, and in thermal equilibrium with a heat bath at constant
temperature. The density distribution of the electrolyte ions in the
region $G$ exterior to the colloidal spheres is obtained from the
normalized electrostatic potential $\phi({\bf r})$ satisfying the PB
equation
\begin{equation}
  \label{eq:1}
\begin{array}{rclcl}
 \nabla^{2} \phi({\bf r}) & = & \kappa^{2} \sinh \phi({\bf r})\:, 
& \qquad  &{\bf r} \in G \\
{\bf n}_{i}\nabla \phi & = & 4 \pi \lambda_{B} \sigma& &{\bf r}
\in \partial G_i, \: i=1,\ldots,N \:,
\end{array}
\end{equation}
where $\partial G_i$ is the surface of the $i$-th colloid with outward
pointing surface normal ${\bf n}_{i}$.  Constant-charge boundary
conditions are assumed for all $N$ colloid surfaces \cite{rem1}.  The
PB-MD simulation algorithm is now described as follows
\cite{technical_paper}: (i) fix all $N$ colloids at their initial
positions $\{{\bf R}_i\}^{0}$, (ii) solve the PB eq.~(\ref{eq:1}) for
this colloidal configuration, (iii) calculate the total force on each
colloid by integrating the stress-tensor over a surface enclosing the
respective particle, (iv) move all colloids to their new positions by
an Euler MD time-step including a stochastic force modeling the heat
bath.  Repeat (ii - iv).  The solution of the PB equation in step (ii)
is computationally the most demanding part of the calculation since it
is necessary to resolve the steep variation of $\phi$ near the colloid
surfaces.  To achieve this resolution with a reasonable number of
grid-points, $N$ spherical coordinate systems are constructed, one
centered around each colloid, which overlap with a Cartesian system in
the cubic simulation box.  The solution is obtained by (iia) solving
the PB equation in each of the spherical systems for fixed values at
the outer edges, (iib) interpolating to find the boundary values in
the Cartesian system, (iic) solving the PB equation in the Cartesian
system, and (iid) interpolating back to find new boundary values at
the edges of the spherical systems.  These steps are iterated to
convergence. There are a few studies
\cite{loewen3275,loewen673,tehver1335,fushiki6700} where similar
simulations have been performed, but they mainly concentrate on
determining the structure of colloidal dispersions. The method used by
L\"owen et al. \cite{loewen3275,loewen673} is based on a density
functional approach and includes also the microion correlations.
However, it has been shown in various studies \cite{groot,valleau}
that the microion correlations play only a negligible role in case of
monovalent salt ions as considered in this paper. Groot \cite{groot}
quantified the validity of the PB approach by comparing it to a cell
model Monte-Carlo simulation and showed that the deviations are
already tiny for a ratio $\lambda_{B}/a = 0.03$.  Since we are here
considering an even smaller value $\lambda_{B}/a=0.012$, the PB
approach is perfectly justified relying on Groot's work.
\begin{figure}
\epsfig{file=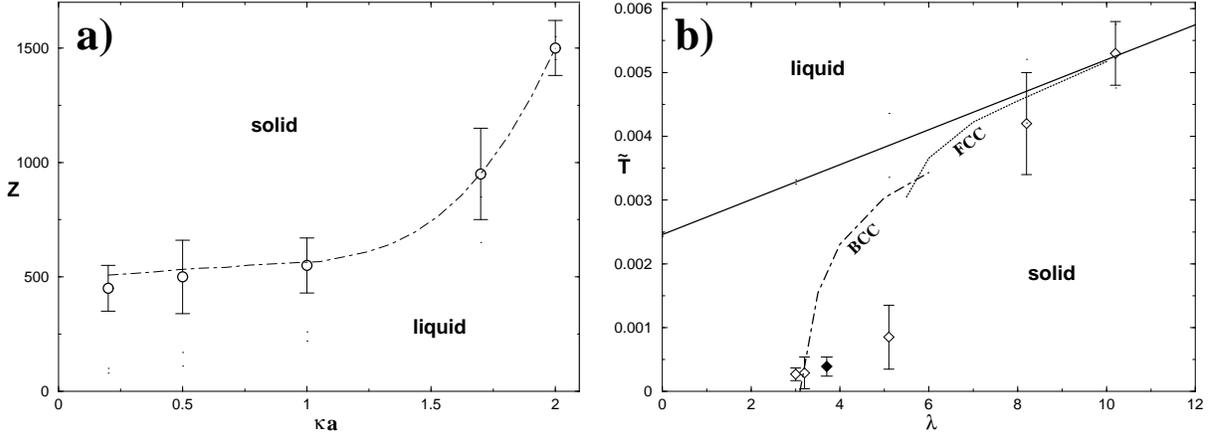,width=\textwidth}
\caption{
  (a) Solid-liquid phase diagram of a charge-stabilized colloidal
  suspension, spanned by the colloidal charge $Z$ and the salt
  concentration $\kappa a$ for volume fraction $\eta=0.03$ and
  $\lambda_{B}/a=0.012$. The circles are points on the melting line
  obtained by applying the Lindemann criterium to PB-MD simulation
  data in a FCC crystal which fully include many-body interactions
  between the colloids.  The line is a guide to the eyes.  (b) The
  melting line of a pure Yukawa system obtained by Robbins et al.
  \cite{rkg} (solid line), the melting line obtained assuming Yukawa
  interactions with a density-dependent cut-off after the first
  neighbor shell in a FCC configuration (dotted line) and in a BCC
  configuration (dot-dashed line), and the data from
  Fig.~(\ref{fig1}.a) transfered by using the effective force
  parameters as described in the text.
}\label{fig1}
\end{figure}

Using the simulation method just described, we have calculated the
melting line in the phase-diagram of charge-stabilized colloidal
suspensions. To facilitate later comparison of our results with the
classic phase-diagram of Yukawa systems calculated by Robbins, Kremer
and Grest (RKG) \cite{rkg}, we followed their work and determined the
phase-boundary by the Lindemann rule which states that a crystal
begins to melt when the rms displacement in the solid phase is a
fraction of 19 \% of $d_{\nn}$. Taking the radius $a$ of the colloidal
spheres as unit-length scale and having in mind an aqueous system at
room temperature where $\lambda_{B}=7.2\:\AA$ and colloid particles of
size $a = 60$nm as common in experiments we fix the ratio
$\lambda_{B}/a = 0.012$. The state space of our colloidal system then
is three-dimensional, spanned by $\kappa a$ (salt concentration), $Z$
(colloidal charge) and $d_{\nn}/a$ (colloid density), or,
equivalently, the volume-fraction $\eta=4 \pi (a/d_{\nn})^{3}/3$.  A
point $\vec{X}$ in state space is thus given by $\vec{X} \equiv
(\kappa a, Z, \eta)$.  Since a systematic exploration of the full 3D
phase-diagram is too expensive computationally, we focus on a 2D cut
at large volume fractions $\eta = 0.03$ realized in experiments.  The
resulting phase-diagram in the $(Z,\kappa a)$-plane shown in
Fig.~(\ref{fig1}.a) represents our main result. It describes the
solid-liquid phase behavior of colloidal suspensions with all
many-body interactions among the colloids included. To our knowledge,
it is the first such phase diagram calculated without assuming
pairwise additive effective interactions. Its significance is best
appreciated by contrasting it to the predictions based on the
established description in terms of pairwise Yukawa potentials, which
in turn requires to first introduce the concepts of charge
renormalization and effective forces.

For a system of {\em point-like} Yukawa particles, interacting via
$u(r) = U_{0} e^{-\lambda r/d_{\nn}}/(r/d_{\nn})$, the state space is
two-dimensional and spanned by $U_{0}$ and $\lambda$. In this space
the melting line is a function $U_{0}^{M}(\lambda)$ that has been
determined by RKG.  More precisely, they introduced an effective
temperature $\tilde{T}$ ($kT$ in units of the Einstein phonon energy)
which is related to $U_{0}$ by $\tilde{T}(\lambda) = (2 \lambda^{2}
\theta(\lambda) \beta U_{0}(\lambda) /3)^{-1}$ where the function
$\theta(\lambda)$ is given in Tab.~I of \cite{rkg}.  They then determined
the melting line as $\tilde{T}^{M}(\lambda) = 0.00246 + 0.000274
\lambda$ which is plotted as the solid line in Fig.~(\ref{fig1}.b).
For colloidal systems, we recall that the Yukawa pair-potential of
DLVO theory reads
\begin{equation}
  \label{eq:2}
\beta u(r) = \Big[\frac{Z_{\eff} e^{\kappa_{\eff} a}}{1+\kappa_{\eff} a}\Big]^{2}
\lambda_{B} \frac{e^{-\kappa_{\eff} r}}{r}
\end{equation}
with the effective colloidal charge $Z_{\eff}$ and the effective
screening parameter $\kappa_{\eff}$. These effective (or renormalized)
quantities, introduced to capture effects arising from the
non-linearity of the PB equation \cite{belloni_review}, are both
(unknown) functions of $\vec{X}$, $\kappa_{\eff} a = f_{1}(\vec{X})$
and $Z_{\eff} = f_{2}(\vec{X})$. One way, among others, to
approximately determine these functions is to use the PB cell model
\cite{alexander,trizac}.  Identifying the parameters in
eq.~(\ref{eq:2}) with $U_{0}$ and $\lambda$ from the RKG calculation
as $\beta U_{0} = (Z_{\eff} e^{\kappa_{\eff} a}/(1+\kappa_{\eff}
a))^{2}\lambda_{B}/d_{\nn}$ and $\lambda = \kappa_{\eff} d_{\nn}$, one
can then use the RKG phase-diagram in Fig.~(\ref{fig1}.b) to estimate
whether a colloidal system, represented by a state point $\vec{X}$, is
solid or liquid. Note first that since the phase-behavior of a pure
Yukawa system is determined just by $U_{0}$ and $\lambda$, the natural
3D state space of colloidal systems is reduced here to 2D.  Secondly,
it is clear that the data in Fig.~(\ref{fig1}.a) can be transfered to
the phase-diagram of Fig.~(\ref{fig1}.b), only if the two functions
$f_{1}(\vec{X})$ and $f_{2}(\vec{X})$ are provided.  Then a data point
at $(Z,\kappa a)$ in Fig.~(\ref{fig1}.a) can be transformed to
$(Z_{\eff},\kappa_{\eff} a)$ and subsequently plotted as
$(\tilde{T},\lambda)$ in Fig.~(\ref{fig1}.b). This requires two steps:
(i) to calculate the effective pair-interaction (depending on the
state $\vec{X}$), which includes the original many-body interactions
and, in general, is not a Yukawa potential and then try (ii) to fit
these pair-interactions to Yukawa potentials.
\begin{figure}
\epsfig{file=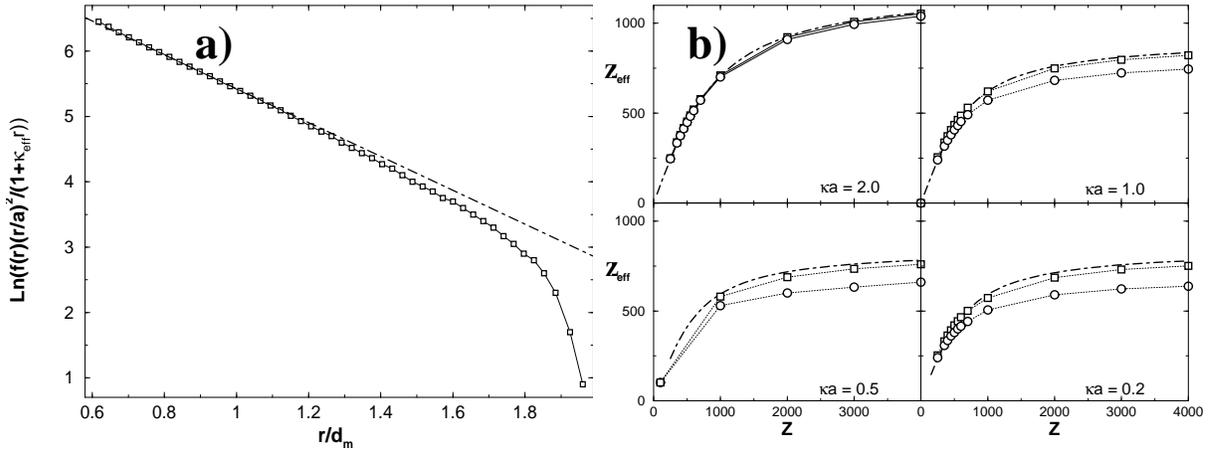,width=\textwidth}
\caption{
  (a) Effective colloid-colloid pair-force as function of the distance
  between two colloids surrounded by 106 other colloids in a FCC
  configuration at $\eta =0.03$, $Z=1000$, $\kappa a=0.2$,
  $\lambda_{B}/a=0.012$. The dot-dashed line is the best fitting
  Yukawa interaction. (b) $Z_{\eff}$ vs. $Z$, obtained from fitting
  Yukawa pair-forces to effective force curves in BCC (squares) and
  FCC (circles) configurations, for four salt concentrations as
  indicated and $\eta =0.03$, $\lambda_{B}/a=0.012$.  The dot-dashed
  lines gives predictions of the PB cell model \cite{alexander} for
  comparison.}
\label{fig2}
\end{figure}

We proceed with the first task of finding an effective
pair-interaction by performing step (i) to (iii) of the simulation
algorithm described above for fixed colloidal configurations.  The
force $\vec{F}_{AB}$ exerted by a particle A on another particle B is
obtained as the difference between the total force acting on particle
B with particle A present, $\vec{F}_{B}^{1}$, and the total force
acting on particle B, $\vec{F}_{B}^{0}$, after removing particle A
while leaving all other particles at their positions
\cite{belloni_idee}. Varying the position of particle A results in the
effective force curve $\vec{F}_{AB}(\vec{r})=
\vec{F}_{B}^{1}(\vec{r})-\vec{F}_{B}^{0}$.  With this procedure the
many-body interactions are folded into an effective pair interaction.
If the true interactions in the system are pairwise additive, the
resulting effective interaction is by construction identical to the
true pairwise interaction potential, \emph{independent} of the
arrangement of the surrounding particles (all particles but A and B).
This calculation has been carried out for FCC ($N = 32$ and 108) and
BCC ($N=54$) configurations at different salt concentrations between
$\kappa a = 0.1$ and $\kappa a = 2.0$, bare colloidal charges between
$Z = 10$ and $Z = 4000$, and several volume fractions $\eta = 0.01,
0.02, 0.03$. A typical effective force curve $\vec{f}(r) = a \beta
\vec{F}_{AB}(r)$ in a low-salt system ($\kappa a=0.2$) is presented in
Fig.~(\ref{fig2}.a). Forces are multiplied by $(r/a)^{2}
/(1+\kappa_{\eff} r)$ and plotted logarithmically so that the Yukawa
interaction appears as a straight line with slope $-\kappa_{\eff}$.
Obviously, for small distances the effective forces are Yukawa-like.
However, around the mean colloid-colloid distance $d_{\nn}$ systematic
deviations become visible which develop into a cut-off at $r>d_{\nn}$.
At the highest salt concentrations considered ($\kappa a=2.0$), no
such deviation occur in our calculation. A similar behavior has also been
found in primitive model calculations \cite{klein}.

As a second step, the values of the effective parameters $Z_{\eff}$
and $\kappa_{\eff}$ are obtained by fitting the force curve to the
derivative of eq.~(\ref{eq:2}) at $r\le d_{\nn}$, that is, in the
range where the force is Yukawa-like. Performing the fit for all
calculated force curves yields the dependence of the effective
parameters on the state of the system, i.e., the two functions
$f_{1}(\vec{X})$ and $f_{2}(\vec{X})$.  The dependence of $Z_{\eff}$
on $Z$ for different salt concentrations is shown in
Fig.~(\ref{fig2}.b). The values obtained in FCC and BCC configuration
are compared to the curves obtained from the PB cell model using the
Alexander prescription \cite{alexander,trizac}. The predictions of the
cell model agree with our results in the high-salt case ($\kappa a =
2.0$), but not under low salt conditions ($\kappa a = 0.2$).
Moreover, we find agreement between the FCC and BCC results at $\kappa
a = 2.0$ but substantial differences at $\kappa a = 0.2$.  As noted
above, such a configuration dependence
($f_{1,2}^{\mathrm{BCC}}(\vec{X})\ne f_{1,2}^{\mathrm{FCC}}(\vec{X})$)
of the effective pair-interaction parameters is another fingerprint of
the many-body interactions in the system next to the observed cut-off
feature.  A dependence of the pair-interaction on the colloidal
configurations has also been found in \cite{loewen673}. In addition,
attractive three-body forces have been explicitly shown to become
effective in the low salt regime ($\kappa a < 1$) \cite{carsten}.  The
range of the effective interaction in the high salt calculation
($\kappa a = 2.0$), is about $1/\kappa_{\eff}\approx 0.5 a$, while for
the low salt case ($\kappa a = 0.2$) it is roughly $2a$. The mean
distance $d_{\nn}$ in both cases is about $5a$. The onset of the
configuration dependence thus coincides with the size of the double
layer becoming comparable to $d_{\nn}$ as pointed out in the
introduction.

Having determined the functions $f_{1}(\vec{X})$ and $f_{2}(\vec{X})$
the data for the melting line in Fig.~(\ref{fig1}.a) can be transfered
to Fig.~(\ref{fig1}.b), essentially by rescaling twice the x- and
y-axis of Fig.~(\ref{fig1}.a) ($(Z,\kappa a) \to
(Z_{\eff},\kappa_{\eff} a) \to (\tilde{T},\lambda)$).  Good agreement
with RKG is obtained in our high-salt calculation $\kappa a = 2.0$,
corresponding to large values of $\lambda$, which is consistent with
our finding that at high salt neither a cut-off behavior nor a
configuration-dependence of the pair-potential could be observed.
Obviously, in this salt regime the colloidal suspension can be
represented by a Yukawa system quite well. However, reducing the
amount of salt, i.e., decreasing $\lambda$, we observe pronounced
deviations from the RKG line occurring in the low salt regime. Again,
this matches with the behavior of the calculated effective force
curves at low salt, showing a configuration dependence but also the
cut-off feature. Both observations suggest that the difference between
the RKG melting line and ours is due to many-body effects.

For practical applications it would of course be highly desirable to have
a way of predicting the phase behavior of colloidal systems 
from a simple density-dependent pair-potential. 
The cut-off behavior observed in 
our effective force curves at low salt concentration and in 
the experiment in \cite{brunner}
suggests to use a truncated Yukawa potential 
\begin{equation}
\label{eq:3}
u(r) = \left\{ \begin{array}{ll}
U_{0} \frac{e^{-\lambda r/d_{\nn}}}{(r/d_{\nn})} & r\le r_{c}\\
0 & r > 0 \end{array} \right.
\end{equation}
with a {\em density-dependent} cut-off $r_{c} \propto d_{\nn}$ to
include the macro-ion shielding effect \cite{klein}. With this model
potential, we have carried out MD simulations and again determined the
solid-liquid phase-boundary by the Lindemann criterion, computing the
rms displacement for various combinations of $U_{0}$ and $\lambda$ in
FCC and BCC crystals. For $r_{c} = 3.07 d_{\nn}$, the cut-off used for
numerical reasons in \cite{rkg}, the RKG melting line is reproduced.
Upon decreasing the cut-off, systematic deviations from the RKG
melting line are observed occurring first at small values of
$\lambda$. The dotted line in Fig.~(\ref{fig1}.b) is the melting line
obtained choosing $r_{c} = 1.35 d_{\nn}$ in a FCC configuration, while
the dash-dotted line is the melting line in a BCC configuration
choosing the cut-off radius $r_{c}=1.50 d_{\nn}$. These two cut-off
radii correspond to the first neighbor shell in the FCC and to the
second neighbor shell in the BCC configuration so that the number of
neighbors that are included in the interaction is comparable in both
cases.  By comparing these results to our full PB-MD simulation it is
seen that the density-dependent cut-off provides a considerable
improvement over the plain Yukawa potential.  Since the cut-off in
this model pair-interaction is related in a simple way to the
macro-ion shielding effect this further corroborates our conclusions
about the importance of many-body interactions for the solid-liquid
phase behavior of colloidal suspensions.

Considering three-body forces for three colloids in an electrolyte,
\cite{carsten}, one can show that the macroion shielding effect
becomes less efficient with decreasing $a$. This implies that the
observed difference between the RKG line and our line should become
smaller with decreasing $\eta=4 \pi (a/d_{\nn})^{3}/3$, which indeed
we have observed in reference calculations at smaller volume fractions
($\eta=0.01$ and $\eta=0.005$). This is, first of all, a reminder of
the fact, already discussed above, that the phase-diagram of colloidal
suspension depends on three quantities, while the Yukawa phase-diagram
is just a 2D representation of it. Secondly, this is the reason why
the existing experimental data in \cite{Gast,Palberg} can not be used
to validate our results, as these experiments have been performed at
even smaller volume fractions ($10^{-3}$ to $10^{-4}$). For such low
volume fractions we do not expect noticeable many-body effects. We
would therefore suggest to perform experiments similar to the ones in
\cite{Gast,Palberg} at higher volume fractions comparable to
$\eta=0.03$ considered in our simulation.

A charge-stabilized colloidal suspension is often considered as a
simple Yukawa liquid. This is not always correct. In conclusion, this
paper shows that a Yukawa description of colloidal suspensions fails
to predict the correct solid-liquid phase behavior under low-salt
conditions where the range of the effective interaction is comparable
to the mean distance and where many-body effects start playing a vital
role.  We have demonstrated this with the help of effective force
calculations which have revealed a configuration dependence of the
pair-interactions and deviations from Yukawa behavior, at $r\ge d_{m}$
and low salt. Model pair-potentials with density-dependent cut-offs
have been shown to reproduce quite well the effects which many-body
interactions have on the phase-behavior of colloidal suspension. We
predict many body effects to increase with increasing colloid volume
fraction. 

We gratefully acknowledge useful and stimulating discussions with
R. Klein, C. Bechinger, M. Brunner, Y. Chen, M. Brumen, D. Halo\v zan and C Russ.


\begin{thebibliography}{99}
\bibitem{rem1}
  Integrating the microioinic density distribution over $G$ results in
  the total number of microions in the system, which allows to compute
  $n_{s}$, the concentration of salt in the system.  The latter is to
  be distinguished from the reservoir salt concentration $c_{s}$ (due
  to the Donnan effect $n_{s}\le c_{s}$).
\end{thebibliography}
\end{document}